# Isospin and Isospin/Strangeness Correlations in Relativistic Heavy Ion Collisions

Aram Mekjian

Rutgers University, Department of Physics and Astronomy, Piscataway, NJ. 08854
&
California Institute of Technology, Kellogg Radiation Lab 106-38, Pasadena, Ca 91125

## Abstract

A fundamental symmetry of nuclear and particle physics is isospin whose third component is the Gell-Mann/Nishijima expression $I_Z = Q - (B+S)/2$. The role of isospin symmetry in relativistic heavy ion collisions is studied. An isospin $I_Z$, strangeness $S$ correlation is shown to be a direct and simple measure of flavor correlations, vanishing in a $Qg$ phase of uncorrelated flavors in both symmetric $N = Z$ and asymmetric $N \neq Z$ systems. By contrast, in a hadron phase, a $I_Z / S$ correlation exists as long as the electrostatic charge chemical potential $\mu_Q \neq 0$ as in $N \neq Z$ asymmetric systems. A parallel is drawn with a Zeeman effect which breaks a spin degeneracy

PACS numbers: 25.75.-q, 25.75.Gz, 25.75.Nq

## Introduction

A goal of relativistic high energy collisions such as those done at CERN or BNL RHIC is the creation of a new state of matter known as the quark gluon plasma. This phase is produced in the initial stages of a collision where a high density $\rho$ and temperature $T$ are produced. A heavy ion collision then proceeds through a subsequent expansion to lower $\rho$ and $T$ where the colored quarks and anti-quarks form isolated colorless objects which are the well known particles whose properties are tabulated in ref[1]. Isospin plays an important role in the classification of these particles [1,2]. A discussion of its properties and consequences in relativistic heavy ion collisions seems useful. Isospin has also been used in the study of medium energy heavy ion collisions[3]. Ref.[3] contains a series of reprints illustrating its importance. In low energy nuclear physics, the isospin symmetries have proven to be very useful in the classification of nuclear levels. Isospin symmetries are broken by Coulomb interactions. Part of this paper explores a parallel with this multiplet structure and it's breaking by Coulomb interactions where the Coulomb interaction is now replaced with the electrostatic $\mu_Q$. Systems with large isospin excess will be explored in future RIA experiments and at FAIR [4]. The extension of isospin symmetry for strongly interacting particles to the weak sector involves weak isospin $I_Z^W$ which is a fundamental symmetry of the standard model with very important consequences [2]. $I_Z^W$ is also given by a Gell-Mann/Nishijima expression.

For nuclei and particles made of $u, d$ quarks only, the third component of isospin $I_Z$, charge $Q$ and baryon number $B$ are related by $2I_Z = 2Q - B$. For u, d, and s quarks, strangeness is incorporated into the connection between isospin and $Q, B$ through the use of hypercharge $Y = B + S$ (+charm $C$ +bottom $\hat{B}$ + top $\hat{T}$; only $B, S$ will be considered). The Gell-Mann/Nishijima [2] equation is $2I_z = 2Q - Y$, a generalization of the previous expression $2I_z = 2Q - B$. In the $Qg$ phase $2I_z = U - D = (N_u - N_{\bar{u}}) - (N_d - N_{\bar{d}})$, where $N_j$ ($j = u, d, \bar{u}, \bar{d}$) are the number of up, down quarks and anti-quarks. In a flavor unlocked or flavor uncorrelated Qg phase the correlation of isospin $I_z$ with $S = N_{\bar{S}} - N_S$, and also charm $C$, top $\hat{T}$, bottom $\hat{B}$, would vanish for any symmetric $N = Z$ or asymmetric $N \neq Z$ system with neutron number $N$ and proton number $Z$. This may not be the case in color flavor locked CFL phases [5] which may occur at high baryon chemical potential $\mu_B$ and low $T$. Here the study centers on the parts of phase space away from this region and in particular regions explored by experiments such as those at RHIC/CERN. However, even in these regions, the chromoelectric plasma may have a more complicated structure than an ideal plasma of quarks and gluons [6,7]. A $I_z / S$ correlation can be used as a measure of flavor correlations in a non-ideal plasma and in CFL phases.

As a baseline for comparison the assumption will be made of uncorrelated flavors in the Qg phase and the question that is addressed is: how large is this correlation in a hadron phase? To answer this question a statistical model in a grand canonical form will be used. A grand canonical statistical approach has been shown to be a useful description of the hadron phase [8-10]. Previous studies [11,12,13,14] showed the importance of looking at fluctuations and correlations as a probe of a phase transition.

A parallel can also be drawn with a Zeeman splitting of different spin $J_z$ levels in a multiplet by an external magnetic field. Population of these levels in a system at finite temperature $T$ will differ by the Boltzmann factor in energy $E(J_z)/T$. In this paper an "external ($N - Z$) field" splits the population of the different members of the isospin multiplet according to their $I_z$. This splitting arises from a Boltzmann factor in $\mu_Q / T$ where the charge chemical potential $\mu_Q$ is the systems response to this "external field".

## 2. Statistical Model Analysis in a Hadron Phase
### 2.1 General Results

The statistical model [8-10] assumes that equilibrium in the strongly interacting sector is established in a volume $V$ and at a temperature $T$. In the grand canonical ensemble three chemical potentials appear in the expression for the particle yields $<N_i>$ which comes from the constraints of baryon number, charge and strangeness. The $<N_i>$, in the non-degenerate limit, will be written here in a form which reads

$$<N_i> = a_i x^{b_i} y^{q_i} z^{(-s_i)} \qquad (1)$$

As an example, the particle $\Delta^{++}$ has $<N_{\Delta^{++}}> = a_{\Delta^{++}} x y^2$. The anti-particle of $\Delta^{++}$ has $<N_{\bar{\Delta}^{++}}> = a_{\bar{\Delta}^{++}} x^{-1} y^{-2}$. The simple quark model restricts $b_i = \pm 1, 0$, $q_i = \pm 2, \pm 1, 0$, and $s_i = \pm 3, \pm 2, \pm 1, 0$. The $a_i = g_s(i)(V/2\pi^2)(m_i/T)^2(K_2(m_i/T))$ with $g_s(i) = 2s_i + 1$ and mass $m_i$. The $x \equiv \exp[\mu_B/T]$, $y \equiv \exp[\mu_Q/T]$ and $z \equiv \exp[-\mu_S/T]$. The $x, y, z$ are determined by constraints on net baryon number $B$, net charge $Q$ and strangeness $S$:

$B = N_B - N_{\bar{B}} = \Sigma b_I < N_i >$, $Q = N_{Q+} - N_{Q-} = \Sigma q_i < N_i >$, $S = N_{S+} - N_{S-} = \Sigma s_i < N_i >$.

In a heavy ion collision the net strangeness is zero, but correlations associated with $S$ and some other quantity may not necessarily be zero. As an obvious example, the correlation of $S$ with itself is not $=0$: $<S^2> - <S>^2 = <S^2> \neq 0$ when strange particles are produced. Because total $B, Q, S$ are each conserved so are any combinations of them such as hypercharge $Y = B + S = \Sigma(b_i + s_i) < N_i >$ and $I_Z$ or $2I_Z$ given by the Gell-Mann/Nishijima result $2I_Z = 2Q - Y = \Sigma(2q_i - b_i - s_i) < N_i > = Z - N$. The $Y$ or $2I_Z$ equation can also replace one of the three previous constraint equations to determine the chemical potentials. As will be shown, the $2I_Z$ isospin equation is very useful for determining the electric chemical potential and the hypercharge equation is also useful in some specific situations such as when $\Xi, \Omega$ contributions are small.

To proceed, small mass differences between different charge states of the same particle will be ignored. The $a_i$ is then the same for each $I_Z$ state of a given particle. Differences in $< N_i >$ between different charge states in an isospin multiplet will occur when $y \neq 1$. Contributions of excited states of a given particle can be added to the lowest mass contribution so that $a_i \to A_i$. For instance, the number of $\Sigma$ like particles $<\Sigma(1267)> + <\Sigma^*(1378)> + \ldots = A_\Sigma xz(y + 1 + 1/y)$ where the new coefficient $A_\Sigma = a(\Sigma(1267)) + a(\Sigma^*(1378)) \ldots$. The $A_\pi$ will contain $\rho$ and $A_N$ has $p, n, N^*$. Particles are grouped according to baryon number, strangeness and isospin $\vec{I}$ with $J = \{N, \Delta, \Lambda, \Sigma, \Xi, \Omega, \pi, K\}$. Resonance decays following freeze out are discussed below.

*2.2 Coupled equations for $x, y, z$ or the chemical potentials $\mu_B, \mu_Q, \mu_S$. Connection of $I_Z$ & $\mu_Q$.*

*2.2a General expressions*

The $B = N_B - N_{\bar{B}}$ constraint and $S$ constraint equations in $(x, y, z)$ variables are

$$B = \sum_J b_J A_J x^{b_J} z^{-s_J} \sum_{q_J} y^{q_J}, \quad S = \sum_J s_J A_J x^{b_J} z^{-s_J} \sum_{q_J} y^{q_J} \quad (2)$$

For instance, $\Delta$ would make a contribution $A_\Delta x(y^2 + y + 1 + y^{-1})$. The $2I_Z$ will be used as one of the three equations to determine the three unknowns $x, y, z$. Specifically,

$$\begin{aligned} 2I_Z &= 2I_Z(B) - 2I_Z(\bar{B}) \\ &= A_N x(y-1) + A_\Delta x(3y^2 + y - 1 - 3y^{-1}) + A_\Sigma xz(2y + 0y - 2y^{-1}) + A_\Xi xz^2(-y^{-1} + 1) \\ &\quad + A_\pi(2y + 0y - 2y^{-1}) + A_K\{z(-y^{-1} + 1) + (1/z)(y-1)\} - 2I_Z(\bar{B}) \end{aligned} \quad (3)$$

The $2I_Z(\bar{B})$ is the anti-baryon part of Eq.(3) and is obtained from $2I_Z(B)$ by taking the reciprocal of each $x, y, z$. The isosinglets $\Lambda^0, \Omega^-$ do not contribute, nor do the $I_Z = 0$ component of the isotriplets such as $\Sigma^0, \pi^0, \rho^0$. The coefficients in front of $y$, when divided by 2, are just the isospin $I_Z$ component of each charged state of each particle $J$. The $I_Z$ equation explicitly shows that when $I_Z = 0$ then $y = 1$. Specifically, every term in ( ) parenthesis involving $y$ in Eq.(3), such as $(3y^2 + y - 1 - 3y^{-1})$, is zero at $y = 1$.

At $y = 1$, the $\mu_Q = 0$. This result is true only when the mass differences between members of an isospin multiplet are neglected. Mass splittings such as in $m_p - m_n \approx -1.3 MeV$, $m_{\Sigma^+} - m_{\Sigma^-} \approx 8 MeV$, $m_{K^+} - m_{K^0} \approx -4 MeV$ give $\mu_Q \neq 0$ even when $2I_Z = Z - N = 0$. At low $T$ the rhs of Eq.(3) is dominated by $n, p, \Delta, \pi^\pm$. The $\pi^+$ and $\pi^-$ have the same mass, and $\delta m$ in the $\Delta$ multiplet is not known. Unfortunately, the mass splitting contribution from the $\Delta$ isospin multiplet could be the largest-see the factor of 10 in Eq.(4) below which arises from spin and isospin effects. Thus an estimate of $\delta m$ contributions would be unreliable. The $\mu_Q = m_p - m_n \approx -1.3 MeV$ for the $(n, p)$ multiplet at $I_Z = 0$ is small compared to results given below as in Eq.25 with $(N - Z)/B \sim 0.2$ and $T \sim 120 MeV$. For $S = 0$, $2I_Z = Z - N$ and $I_Z = 0$ for symmetric $N = Z$ systems. When $N \neq Z$, $y \neq 1$ and then the various $2I_Z$ charge states of a given multiplet have different $< N_i >$. Usually $\mu_Q << T$. As a result a lowest order expansion around $y = 1$ will be made and only linear terms in $\mu_Q / T$ will be kept. This consequence is a statement of a "weak external field" which leads to a simplifying linear relation $\mu_Q \sim Z - N$. Namely, expanding $y = \exp(\mu_Q / T) = 1 + \mu_Q / T$ in Eq.(3) gives

$$\frac{\mu_Q}{T} = \frac{Z - N}{(A_N + 10 A_\Delta)(x + \frac{1}{x}) + 4 A_\Sigma (xz + \frac{1}{xz}) + A_\Xi (xz^2 + \frac{1}{xz^2}) + 4 A_\pi + A_K (z + \frac{1}{z})} \qquad (4)$$

The numerical coefficients in front of each $A_J$ are given by $\Sigma (2 q_i - (b_i + s_i)) q_i$ with $\Delta^{++}, \Delta^+, \Delta^0, \Delta^-$ contributing 6,1,0,3, respectively. The large factor of 10 makes $\Delta$ important in $\mu_Q / T$. In order to obtain $\mu_Q$, $x$ and $z$ also have to be determined. The evaluation of $x, z$ for $I_Z = 0$ is given below. As a first approximation these values also apply when $I_Z \neq 0$ since $\mu_Q$ is small compared to $\mu_S, \mu_B$. Typical values for a temperature range $T \sim 100 MeV \rightarrow 170 MeV$ have $\mu_S \sim \mu_B / 5$ and for $(N - Z) \sim 0.2 B$ $\mu_Q \sim -\mu_B / 40$. The $\mu_B$ has a behavior given in ref.[9] as $\mu_B = 80.85 (T_0 - T)^{1/2} MeV$, $T_0 = 167 MeV$ by fitting the statistical model to various data. This type of behavior of $\mu_B$ versus $T$ can also arise in a Hagedorn model [15] with a density of excited states $\rho = D_\tau m^{-\tau} \exp(\beta_h m)$, where $D_\tau$ is a constant and $\beta_h = 1/T_0$. The $A_J \sim V (y/m_0)^{\tau - 5/2} (\int_y^\infty (dx e^{-x} / x^{(\tau - 3/2)}))$ with $y = (T_0 - T) m_0 / T_0 T$. The $m_0$ is the lowest mass of particles of type $J$. For $\tau < 5/2$, $A_J \rightarrow \infty$ as $1/(T_0 - T)^{(5/2 - \tau)}$. Moreover, $\mu_B \rightarrow 0$ gives rise to a singular behavior in the heat capacity at $T_0$ [15]. See ref.[15,16] for further consequences of a Hagedorn model.

2.2b *Constraint equations at y=1.*

When $y = 1$ the constraint equations are as follows. The result for $B$ is:

$$B = (2A_N + 4A_\Lambda)(x - \frac{1}{x}) + (A_\Lambda + 3A_\Sigma)(xz - \frac{1}{xz}) + 2A_\Xi(xz^2 - \frac{1}{xz^2}) + A_\Omega(xz^3 - \frac{1}{xz^3}) \qquad (5)$$

The equation for $-S$ is:

$$-S = (A_\Lambda + 3A_\Sigma)(xz - \frac{1}{xz}) + 4A_\Xi(xz^2 - \frac{1}{xz^2}) + 3A_\Omega(xz^3 - \frac{1}{xz^3}) + 2A_K(z - \frac{1}{z}) \qquad (6)$$

The hypercharge $Y = B + S$ equation would read:

$$Y = (2A_N + 4A_\Lambda)(x - \frac{1}{x}) - 2A_\Xi(xz^2 - \frac{1}{xz^2}) - 2A_\Omega(xz^3 - \frac{1}{xz^3}) - 2A_K(z - \frac{1}{z}) \qquad (7)$$

The unknowns $x, z$ can be obtained by solving two of the three Eq. (5,6,7) or any other combinations of them. When $\Xi, \Omega$ can be neglected ($A_\Xi, A_\Omega$ are small compared to other terms) these equations simply considerably since higher order powers of $z$ ($z^2, z^3, 1/z^2, 1/z^3$) are absent. Such solutions will be given below.

2.3 *Isospin correlations in the Qg phase and hadron phase.*
In the $Qg$ phase the charge $Q$, baryon number $B$, strangeness $S$, hypercharge $Y$, isospin $I_Z$ and a quantity $L_Z \equiv 2Q - B$ are given by

$$Q = \frac{2}{3}(N_u - N_{\bar{u}}) - \frac{1}{3}(N_d - N_{\bar{d}}) - \frac{1}{3}(N_s - N_{\bar{s}}) \equiv \frac{2}{3}U - \frac{1}{3}D + \frac{1}{3}S \qquad (8)$$

$$B = \frac{1}{3}(N_u - N_{\bar{u}}) + \frac{1}{3}(N_d - N_{\bar{d}}) + \frac{1}{3}(N_s - N_{\bar{s}}) \equiv \frac{1}{3}U + \frac{1}{3}D + \frac{1}{3}S \qquad (9)$$

$$S = -(N_s - N_{\bar{s}}) \qquad (10)$$

$$Y = B + S = \frac{1}{3}(N_u - N_{\bar{u}}) + \frac{1}{3}(N_d - N_{\bar{d}}) - \frac{2}{3}(N_s - N_{\bar{s}}) \equiv \frac{1}{3}U + \frac{1}{3}D + \frac{2}{3}S \qquad (11)$$

$$2I_Z = 2Q - Y = (N_u - N_{\bar{u}}) - (N_d - N_{\bar{d}}) \equiv U - D \qquad (12)$$

$$L_Z \equiv 2Q - B = (N_u - N_{\bar{u}}) - (N_d - N_{\bar{d}}) - (N_s - N_{\bar{s}}) \equiv U - D + S \qquad (13)$$

As already mentioned, if flavors are uncorrelated, the correlation between isospin and strangeness (and also charm $C$, topness $\mathcal{T}$ and bottomness $\mathcal{B}$) is zero since $I_Z$ only involves up and down quarks and anti-quarks. Then $<2I_Z X> - <2I_Z><X> = 0$ with $X = S, C, \mathcal{T}$ or $\mathcal{B}$. The $I_Z / S$ correlation is thus a direct, simple measure of flavor correlations. Besides an $I_Z / S$ correlation discussed here, $I_Z$ correlations with baryon number $B$, charge $Q$, hypercharge $Y$ differ in the $Qg$ and hadron phase because the fundamental units of baryon number are ±1/3 and charge are ±1/3e,±2/3e in the *Qg* phase. This property was first recognized for charge fluctuations $<Q^2> - <Q>^2$ [11,12].. The isospin/baryon number correlation is determined by the result $<2I_zB> = 1/3<U^2> - 1/3<D^2>$ with $<B> = 1/3<U> + 1/3<D>$ since $<S> = 0$, $<2I_z> = <U> - <D>$ so that $<2I_z><B> = (1/3)(<U>^2 - <D>^2)$. Thus $<2I_zB> - <2I_z><B> = 1/3(<U^2> - <U>^2) - 1/3(<D^2> - <D>^2)$ vanishes only if $N = Z$. The $<2I_zY> - <2I_z><Y> = 1/3(<U^2> - <U>^2) - 1/3(<D^2> - <D>^2)$. The $<2I_zQ> - <2I_z><Q> = 2/3(<U^2> - <U>^2) + 1/3(<D^2> - <D>^2)$ doesn't vanish even in $N = Z$ systems. When $N = Z$, $<D^2> = <U^2>$ and $<D> = <U>$. The study of such correlations may give some insight into more complicated views of the Qg system [6,7].

In a hadron phase the $I_z/S$ correlation is given by isospin splitting yields:

$$-(<2I_zS> - <2I_z><S>) = -<2I_zS> = \qquad (14)$$

$$2<(\Sigma^+ - \Sigma^-) + (\bar{\Sigma}^+ - \bar{\Sigma}^-)> + 2<(\Xi^0 - \Xi^-) + (\bar{\Xi}^0 - \bar{\Xi}^-)> + <(\bar{K}^0 - K^-) + (K^0 - K^+)>$$

where each $<J>$ can include the lowest mass plus all excited states $J^*$ of that $J$. Particles like $\Lambda(1520) \to N + K$ don't contribute directly to eq.(14) or indirectly through their resonance decay products by isospin conservation. Isospin conservation gives equal numbers of $\bar{K}^0$ and $K^-$ in the two branches $p + K^-$ and $n + \bar{K}^0$ since the Clebsch-Gordon coefficients are $\pm 1/\sqrt{2}$:

$$/\Lambda(1520), I = 0, I_z = 0> = (1/\sqrt{2})/p>/K^-> - (1/\sqrt{2})/n>/\bar{K}^0> \qquad (15)$$

Similar results also apply to decay of the $\phi$ meson. When $I_z = 0$, $y = 1$ and $<I_zS> = 0$. When $I_z \neq 0$, $<I_zS> \neq 0$. Letting $y = 1 + \mu_Q/T$, then

$$-<2I_zS> \approx (\mu_Q/T)[4(<\Sigma^+ + \bar{\Sigma}^+>) + 2(<\Xi^- + \bar{\Xi}^->) + <K^-> - <K^+>] \qquad (16)$$

The $\mu_Q$ is given by Eq.(4). The value of $-<2I_{W,z}S>$ depends on $x, z$. First the $I_z = 0$ $y = 1$ case will be considered to obtain $x, z$. The $I_z \neq 0$ can be built on this solution by successive iterations.

*2.4 Simplified solutions in symmetric N=Z systems where isospin $I_z = 0$, y=1.*
2.4a *Behavior when $I_z = 0$ and $\mu_S/T \ll 1$ or $z \approx 1$.*

When $\mu_S/T \ll 1$, then $z = \exp(-\mu_S/T) \approx 1$. Substituting $z = 1 - \mu_S/T$ in Eq.(6) gives

$$\frac{\mu_S}{T} = \sinh(\mu_B/T)\frac{1(A_\Lambda + (3A_\Sigma)) + \{2(2A_\Xi) + 3(1A_\Omega)\}}{1^2(2A_K) + \cosh(\mu_B/T)[1^2(A_\Lambda + (3A_\Sigma)) + \{2^2(2A_\Xi) + 3^2(1A_\Omega)\}]} \qquad (17)$$

When $\mu_B/T \ll 1$, $\cosh(\mu_B/T) \to 1$, $\sinh(\mu_B/T) \to \mu_B/T$, which can be substituted into Eq.(17) to give a simple linear connection between $\mu_S/T$ and $\mu_B/T$ which is

$$\mu_S/T = (\mu_B/T)\frac{1(A_\Lambda + (3A_\Sigma)) + \{2(2A_\Xi) + 3(1A_\Omega)\}}{1^2(2A_K) + [1^2(A_\Lambda + (3A_\Sigma)) + \{2^2(2A_\Xi) + 3^2(1A_\Omega)\}]} \quad (18)$$

The numerator in this equation is just the strangeness in hyperons while the denominator is the strangeness squared in each hyperon and in $\bar{K}^0$ and $K^-$.

For large $\mu_B/T$, $\sinh(\mu_B/T), \cosh(\mu_B/T) \to \exp(\mu_B/T)/2$. Substituted this result into Eq.(7) gives a non-linear relation between $\mu_S/T$ and $\mu_B/T$ that is

$$\mu_S/T = \exp(\mu_B/T)\frac{1(A_\Lambda + (3A_\Sigma)) + \{2(2A_\Xi) + 3(1A_\Omega)\}}{1^2(2A_K) + \exp(\mu_B/T)[1^2(A_\Lambda + (3A_\Sigma)) + \{2^2(2A_\Xi) + 3^2(1A_\Omega)\}]} \quad (19)$$

The $\mu_B/T$ can be obtained from:

$$\frac{B}{2} = \sinh(\frac{\mu_B}{T})(2A_N + 4A_\Delta + \frac{2A_K F_{S,0} + (F_{S,0}F_{S,2} - (F_{S,1})^2)\cosh(\mu_B/T)}{2A_K + F_{S,2}\cosh(\mu_B/T)}) \quad (20)$$

where $F_{S,j} = 1^j(1A_\Lambda) + 1^j(3A_\Sigma) + 2^j(2A_\Xi) + 3^j(1A_\Omega)$. Since each $A_J \sim V$, $\mu_B$ is a function of $B/V, T$. When $x \gg 1$, and $z$ is small, the main contribution comes from $S = \pm 1$ strange particles ($K, \Lambda, \Sigma$). In general, the baryon constraint equation is quadratic in x and involves only $z$ for $y = 1$. This quadratic equation for $x$ can be easily solved, giving $x = x(z)$ which can be substituted into $S = 0$, to find $z$ straightforwardly.

2.4b Behavior when $I_Z = 0$ and $A_\Xi$ and $A_\Omega$ are neglected

When $\Xi, \Omega$ are neglected, the relation between $\mu_S/T, \mu_B/T$ is now

$$\tanh(\mu_S/T) = \sinh(\mu_B/T)\frac{1(A_\Lambda + (3A_\Sigma))}{1^2(2A_K) + \cosh(\mu_B/T)[1^2(A_\Lambda + (3A_\Sigma))]} \quad (21)$$

Using $Y = (2A_N + 4A_\Delta)2\sinh(\mu_B/T) + 2A_K 2\sinh(\mu_S/T)$ results in:

$$\frac{B}{2} = \sinh(\frac{\mu_B}{T})(2A_N + 4A_\Delta + \frac{2A_K(A_\Lambda + 3A_\Sigma)}{(2A_K + A_\Lambda + 3A_\Sigma)\sqrt{1 + \frac{8A_K(A_\Lambda + 3A_\Sigma)}{(2A_K + A_\Lambda + 3A_\Sigma)^2}\sinh^2(\frac{\mu_B}{2T})}}) \quad (22)$$

For large $\mu_B/T$, $\cosh(\mu_B/T)$ and $\sinh(\mu_B/T) \to \exp(\mu_B/T)/2$ and Eq.(21) is

$$\exp(2\mu_S/T) = \frac{1}{z^2} \approx 1 + \exp(\mu_B/T) \frac{2m_\Lambda^{3/2} e^{-m_\Lambda/T} + 6m_\Sigma^{3/2} e^{-m_\Sigma/T}}{2m_K^{3/2} e^{-m_K/T}} \tag{23}$$

When $y = 1$, $K^+/K^- = \exp(2\mu_S/T)$. At low $T$ where the $\Sigma$ term is small compared to the $\Lambda$ contribution and where the 1 term is small also, the $K^+/K^-$ ratio is determined by

$$K^+/K^- = \exp(\mu_B/T) \frac{2m_\Lambda^{3/2} e^{-m_\Lambda/T}}{2m_K^{3/2} e^{-m_K/T}} = \frac{m_\Lambda^{3/2}}{m_K^{3/2}} \exp((\mu_B - m_\Lambda + m_K)/T) \tag{24}$$

The $K^+/K^-$ ratio can also be used to discuss the role of dropping masses [17] which leads to an enhancement of this ratio. The ratio involves the exponential factor in $((\mu_B - m_\Lambda + m_K)/T)$ which is sensitive to mass shifts.

2.5 Estimates of $-<2I_z S>$ from an "external field"

Typical collisions at RHIC are $Pb + Pb$ which lead to large neutron excesses amongst the participants with $(N-Z)/B \sim 0.2$. To a good approximation (~1% error at T=120Mev):

$$\mu_Q/T \approx [(Z-N)/B][2 + 4A_\Lambda/A_N]/[1 + 10A_\Lambda/A_N + 4A_\pi/(A_N(x+1/x))]$$
$$\sim [(Z-N)/B] \tag{25}$$

In this approximation the hyperon and strange meson contributions ($A_\Sigma, A_\Xi, A_K$) are small compared to $A_N, A_\Lambda, A_\pi$ in eq.(4) and corresponding, the hyperons are neglected in the baryon constraint equation. Under these conditions $\mu_Q$ has the simple form just given. At $T = 120 MeV$, $\mu_B \sim 600\ MeV$ -see for example the experimental analysis of ref[9] and the theoretical discussion in ref[15] based on a Hagedorn spectrum of states. The associated $\mu_S = 118 MeV$ and z=.3735 from Eq.(21). Including $\Xi, \Omega$ gives $z = .355$ and $\mu_S = 124 MeV$. The $\Xi, \Omega$ lower $z$ and slightly enhance $\mu_S$. At higher temperatures the enhancement from $\Xi, \Omega$ is larger since the population of $\Xi, \Omega$ increases with temperature due to the Boltzmann factors $\exp(-m/T)$. The $\mu_Q = -14.5 MeV$ for $(N-Z)/B = 0.2$. At a higher temperatue $\mu_B, \mu_S$ and the magnitude of $\mu_Q$ decrease. For example at $T = 150 MeV$, $\mu_B \approx 300 MeV$, $z \approx .682$ and $\mu_S \approx 57 MeV$, $\mu_Q \approx -8.5 MeV$. The value of $\mu_Q$ is closely connected to $\mu_B$ and $(N-Z)/B$ through a scaling relation that reads

$$\mu_Q \approx -\frac{N-Z}{B}\frac{\mu_B}{8} \tag{26}$$

The $\mu_S$, $\mu_B$ scaling relation is $\mu_S \approx \mu_B/5$ over a wide range of temperatures that run from $T \approx 100 MeV \to T_0 \approx 167 MeV$. The slope $1/5$ in this linear scaling relation is determined by the strangeness in hyperons to the strangeness fluctuation in strange mesons and hyperons as given by eq.(18) for temperatures $T \approx T_0$.

Once the chemical potentials are determined the $<2I_zS>$ correlation follows from results given in subsect.2.3 The $<2I_zS> \approx 0.018(N-Z)$ at $T=120 MeV$ and $<2I_zS>=0.021(N-Z)$ for $T=150 MeV$. The numerical prefactors, 0.018 for $T=120 MeV$ and 0.021 for $T=150 MeV$ show a slightly increasing temperature dependence. While these coefficients, .018 and .021, are small, the $<2I_zS>$ is obtained by multiplying them by $(N-Z) = 0.2B$ for the example considered here. By contrast, in a flavor uncorrelated $Qg$ phase $<2I_zS>$ is identically=0. A heavy ion collision evolves through an expansion from a high density, high temperature $Qg$ phase to a lower density and lower temperatures hadron phase where the colored quarks and anti-quarks form isolated colorless particles following the QCD phase transition back to the observable hadrons. Flavors become correlated since $<2I_zS> \neq 0$ in the hadron phase for $N \neq Z$ systems. The isospin strangeness correlation in the hadron phase is determined by the electrostatic chemical potential which splits the population of the isospin carrying hyperons $\Sigma, \Xi$ and anti-hyperons $\overline{\Sigma}, \overline{\Xi}$, and strange mesons $K^0, K^+$ and $\overline{K}^0, K^-$. This electrostatic chemical potential breaks the isospin symmetry of the system and produces the correlation $<2I_zS>$.

*2.6 Susceptibilities and lattice calculations*

Correlations and fluctuations can also be related to susceptibilities. Specifically, the $B/S$ correlation defined in [14] as $\hat{C}_{BS} = -3(<BS> - <B><S>)/(<S^2> - <S>^2)$ was also shown to be $\hat{C}_{BS} = -3\chi_{BS}/\chi_{SS}$. The off diagonal baryon/strangeness susceptibility $\chi_{BS}$ and diagonal strangeness susceptibility $\chi_{SS}$ are just derivatives of the grand potential $\Omega$ and in general the relation reads $\chi_{JK} = (-1/V)\partial^2\Omega(V,T,\vec{\mu})/(\partial\mu_j\partial\mu_k)$. In turn, the $\hat{C}_{BS}$ can be related to basic quark flavor susceptibilities of $u,d,s$ [13]. Specifically, using the fact that mean flavor densities vanish at zero chemical potential, the $\hat{C}_{BS} = 1 + (\chi_{us} + \chi_{ds})/\chi_{ss}$. Lattice calculations such as those in ref [18] and ref[19] give some preliminary results on quark susceptibilitie and flavor correlations. In particular, $(\chi_{us} + \chi_{ds})/\chi_{ss}$ =0.00(3)/.53(1) from the results of ref[19,13] showing that flavors are uncorrelated. This result of uncorrelated flavors was assumed in this present work on isospin/ strangeness correlations. The analysis of ref[13] gives a lattice justification for such an assumption. In particular, the analog of $\hat{C}_{BS}$ is

$\hat{C}_{2I_zS} \equiv <2I_zS>/<S^2> = (\chi_{us} - \chi_{ud})/\chi_{ss}$.

**Summary and Conclusions**

This paper explored the role of isospin in relativistic heavy ion collisions using the Gell-Mann/Nishijima formula $2I_z = 2Q - (B+S+C+\hat{B}+\hat{T})$. In particular an isospin-strangeness correlation is discussed. In a quark model of $u,d,s,c,b,t$ quarks, the $2I_z = (N_u - N_{\bar{u}}) - (N_d - N_{\bar{d}})$ and involves only the $u,d$ quarks. The $s,c,b,t$ each has isospin 0. For a heavy ion collision net $S=0$ and net $2I_z = (Z-N)$. In a flavor uncorrelated phase, a correlation between $I_z$ and either $S, C, \hat{B}$ or $\hat{T}$ would vanish. Thus, a non-vanishing $<I_Z S>$ is a simple, direct measure of flavor correlations. With ever improving lattice calculations, the degree of flavor correlations above the QCD transition can be determined. Detailed calculations of some of the consequences of isospin symmetry in the hadron phase were presented using a statistical model. Properties of $I_Z$, $\mu_Q$ and $<I_Z S>$ in this phase were developed. As an example, in a hadron phase an $I_Z / S$ correlation exists when $\mu_Q \neq 0$, even when $S = 0$. An analogy was drawn with a Zeeman splitting of a spin multiplet $(J_Z = -J,..,J)$ in a magnetic field. In this analogy the "external field" is the neutron (or proton) excess $N - Z$ and the Boltzmann factor $\exp[-E(J_Z)/T]$ is replaced with $\exp(\mu_Q/T)$. Expressions are developed, such Eq.(14,16) which can be used to obtain this correlation directly from experimentally determined yields of strange particles. The dependence of $\mu_Q$ on $(N-Z)$ can be determined experimentally by varying $Z$ and $N$. The $I_Z / S$ correlation is a useful method of distinguishing the two phases if a fast expansion of the $Qg$ freezes in the correlations. This fast expansion assumption was the basis for the usefulness of charge fluctuations. Experiments looking for charge fluctuations can be found in ref. [20-23]. This scenario should be contrasted with a situation in which the hadrons form a thermalized medium at the end of the fireball expansion. In this latter situation the thermalized system will have no memory of any prior path. Then information about the $Qg$ phase is lost.

The behavior of $\mu_B, \mu_S, \mu_Q$ is also important for understanding the complete phase diagram of charged, strange, baryon rich hadronic matter. The Gell-Mann/Nishijima formula was shown to be very useful in such a study. Simple analytic expressions or a simple procedure for obtaining these three chemical potentials were given. The $\mu_Q / T$ behavior of Eq.(4) gave a linear relation $\mu_Q \sim (Z-N)$ arising from a response to a "weak external field". The constant between $\mu_Q$ and $(Z-N)$ determines the degree of flavor mixing. This constant involves the isospin doublets like n,p, the isospin triplets such as $\Sigma^+, \Sigma^0, \Sigma^-$ or $\pi^+, \pi^0, \pi^-$ and the isospin quartets - $\Delta^{++}, \Delta^+, \Delta^0, \Delta^-$.

**Acknowledgements**
This work was supported in part by the DOE grant number DE-FG02-96ER-40987.